\documentclass[11pt]{article}
\usepackage{amsmath,amssymb}
\usepackage{ifpdf}

\ifpdf
  \usepackage[pdftex]{graphicx}
\else
  \usepackage[dvipdfm]{graphicx}
\fi

\newcommand{\ba}{\begin{alignat}{3}}
\newcommand{\e}{\epsilon}

\newcommand{\dl}{\delta}
\newcommand{\g}{\gamma}

\newcommand{\tr}{\text{Tr}}

\newcommand{\pa}{\partial}

\newcommand{\si}{\sigma}

\newcommand{\D}{\mathcal{D}}

\newcommand{\lam}{\lambda}
\newcommand{\olam}{\overline{\lambda}}
\newcommand{\oep}{\overline{\epsilon}}

\numberwithin{equation}{section}

\begin{document}

\begin{flushright}
OU-HET 747
\\
April, 2012
\end{flushright}
\vskip1cm
\begin{center}
{\LARGE {\bf Comments on knotted 1/2 BPS Wilson loops}}
\vskip3cm
{\large 
{\bf Akinori Tanaka,\footnote{akinori@het.phys.sci.osaka-u.ac.jp}}

\vskip1cm
\it Department of Physics, Graduate School of Science, 
\\
Osaka University, Toyonaka, Osaka 560-0043, Japan}
\end{center}

\vskip1cm
\begin{abstract}
In this paper, we show that the localization of three-dimensional $\mathcal{N}=2$ supersymmetric Chern-Simons theory on the ellipsoid-like squashed sphere is related to a nontrivial knot structure called torus knot.
More precisely, we can capture the three sphere as the nontrivial so-called Seifert fibrations by regarding 1/2 BPS Wilson loops as $U(1)$ fibers.
The topology of knotted 1/2 BPS Wilson loops is controlled by squashing parameters.
We calculate the 1/2 BPS condition of the Wilson loop and find perfect agreement with known results.
We also remark on the level shift and framing anomaly.
\end{abstract}



\newpage
\tableofcontents


\section{Introduction}\label{intro}
It is well known that the embedding of $S^1$ into $S^3$ makes many nontrivial topological objects such as so-called knots.
This structure is one of the most important ingredients to understand low dimensional topology.
As a physical representation of knot theory, there is a celebrated work by Witten \cite{Witten:1988hf} that proves that the expectation values of knotted Wilson loops weighted by pure Chern-Simons action turn out to be knot invariants.
In particular, when we take the gauge group as $SU(2)$ and the representation of Wilson loop as the fundamental representation, the invariants become the famous Jones polynomials.

On the other hand, in these days, so called localization techniques were developed by Pestun \cite{Pestun:2007rz} in four-dimension, Kapustin, Willett, Yaakov \cite{Kapustin:2009kz} in three-dimension.
Using these techniques, we can calculate exact results of supersymmetric field theories. 
It is mentioned in \cite{Kapustin:2009kz} that $\mathcal{N}=2$ supersymmetric Chern-Simons theory reduces to pure Chern-Simons theory by integrating out gaugino and auxiliary fields. Therefore, it is expected that the expectation value of a supersymmetric Wilson loop becomes knot invariant.

However there is one problem.
All calculable observables using localization technique are $1/2$ BPS ones.
As discussed in \cite{Kapustin:2009kz}, this condition determines the topological structure of Wilson loops completely as trivial knots (unknotted ones).
So one may wonder whether it is possible to modify this localization technique to output nontrivial knots as $1/2$ BPS Wilson loops?

In this paper, we find that one of modification can be accomplished by using Hama, Hosomichi and Lee's localization on ellipsoid-like squashed three spheres \cite{Hama:2011ea}.
In their theory, there are two squashing parameters $l$ and $\tilde{l}$.
We find that if and only if the ratio $l/\tilde{l}$ is a rational number, there exist nontrivially knotted closed $1/2$ BPS Wilson loops.
And we find that these knotted loops construct nontrivial so-called Seifert fibrations, one of the generalizations of the famous Hopf fibration.
Our result matches with known results \cite{Beasley:2009mb, 2011JHEP...08..008K}, and of course matches with Jones polynomials gained by Witten.

The construction of this paper is the followings.
In section 2, we review Hama, Hosomichi, Lee's theory.
In section 3, we discuss the $1/2$ BPS condition.
And in section 4, we comment on some technical details of level shift and the extra phase factors in the results of section 3.

\section{Hama, Hosomichi and Lee's theory}\label{Sec2}
We briefly review the result of Hama, Hosomich and Lee \cite{Hama:2011ea}.
\subsection{Preliminaries}
Let us consider the following ellipsoid metric on $S^3$
\begin{equation}
 ds^2 = f(\theta)^2 d \theta^2 + l^2 \cos^2 \theta d \phi^2 + \tilde{l}^2 \sin^2 \theta d \chi^2, \ \
 f(\theta) = \sqrt{l^2 \sin^2 \theta + \tilde{l}^2 \cos^2 \theta} .
 \end{equation}
 This metric can be regarded as 
\begin{equation}
ds^2 = l^2 (dx_0^2 + dx_1^2) + \tilde{l}^2 (dx_2^2 +dx_3^2),
\end{equation}
where
\begin{equation}
x_0 = \cos \theta \cos \phi, \ x_1 = \cos \theta \sin \phi, \ x_2 = \sin \theta \cos \chi, \ x_3 = \sin \theta \sin \chi . 
\end{equation}
In order to construct supersymmetric theory, we need a Killing spinor.
We must redefine covariant derivatives to maintain the existence of Killing spinors.
Hama, Hosomichi, Lee used usual Killing spinors on round $S^3$
\begin{align}
\e = \frac{1}{\sqrt{2}} 
\begin{pmatrix}
-e^{\frac{i}{2}(\chi - \phi + \theta)} \\
e^{\frac{i}{2}(\chi - \phi - \theta)} 
\end{pmatrix} , \ \ 
\oep = \frac{1}{\sqrt{2}} 
\begin{pmatrix}
e^{\frac{i}{2}(-\chi + \phi + \theta)} \\
e^{\frac{i}{2}(-\chi + \phi - \theta)} 
\end{pmatrix}. \label{killing}
\end{align}
Then we get
\begin{align}
\mathcal{D}_{\mu} \e = \frac{i}{2 f } \g_{\mu} \e, \ \ \mathcal{D}_{\mu} \oep =\frac{i}{2 f } \g_{\mu} \oep . \label{killing2}
\end{align}
where
\begin{align}
\mathcal{D}= d + \frac{1}{4} \omega ^{ab} \g ^{ab}- i q V, \ \ V=V_{\mu} dx^{\mu} = -\frac{1}{2}(1- \frac{l}{f}) d \phi + \frac{1}{2} (1- \frac{\tilde{l}}{f}) d \chi .
\end{align}
They assigned R-charge $q$ as $+1$ to $\e$ and $-1$ to $\oep$,
and to vector multiplets as follows. (See Table 1)
\begin{center}
\begin{tabular}{|c||c|c|c|c|c|} 
\hline
Field & $A_{\mu}$ & $\si$ & $\lam$ & $\olam$ & $D$ \\ \hline \hline
spin & 1 & 0 & 1/2 & 1/2 & 0 \\ \hline
$q$ & 0 & 0 & +1 & -1 & 0 \\ \hline
\end{tabular}

\ \\
Table 1 : Assignments of R-charge $q$
\end{center}
The supersymmetry transformations are
\begin{align}
&\dl _{\e} A_{\mu} = + \frac{i}{2} \olam \g_{\mu} \e  , \ \ \dl _{\oep} A_{\mu} =  - \frac{i}{2} \oep \g_{\mu} \lam ,  \label{Aoep} \\
&\dl _{\e} \si = - \frac{1}{2} \olam  \e  , \ \   \dl _{\oep} \si =  + \frac{1}{2} \oep  \lam ,  \label{sioep} \\
&\dl _{\e} \lam =  \frac{1}{2}  \g^{\mu \nu} \e F_{\mu \nu} - D \e + i \g^{\mu} \e \D_{\mu} \si + \frac{2i}{3} \si \g^{\mu} \D_{\mu} \e, \ \ \dl _{\oep} \lam =  0 ,  \label{lamoep} \\
&\dl _{\e} \olam =  0, \ \ \dl _{\oep} \olam =  \frac{1}{2}  \g^{\mu \nu} \oep F_{\mu \nu} + D \oep - i \g^{\mu} \oep \D_{\mu} \si - \frac{2i}{3} \si \g^{\mu} \D_{\mu} \oep ,  \label{olamoep} \\
&\dl _{\e} D =  - \frac{i}{2}  \D_{\mu} \olam \g^{\mu} \e + \frac{i}{2} [\olam \e , \si ] - \frac{i}{6} \olam \g^{\mu} \D_{\mu} \e, \ \ \dl _{\oep} D = - \frac{i}{2} \oep \g^{\mu} \D_{\mu} \lam + \frac{i}{2} [\oep \lam , \si ] - \frac{i}{6} \D_{\mu} \oep \g^{\mu} \lam  \label{Doep}.
\end{align}
Under these transformations, the supersymmetric Chern-Simons action
\begin{equation}
S_{SCS}(A, \si, \lam, \olam, D) = \int d^3 x \ \sqrt{g} \  \tr \Big[ \frac{1}{\sqrt{g}} \e ^{\mu \nu \lam}(A_{\mu} \pa_{\nu} A_{\lam} - \frac{2i}{3}A_{\mu} A_{\nu} A_{\lam} ) - \olam \lam + 2 D \si \Big]
\end{equation}
is invariant.
And the supersymmetric Yang-Mills action is $\dl_{\oep}$ exact
\begin{align}
S_{YM}(A, \si, \lam, \olam, D) 
&=\int d^3 x \ \sqrt{g} \ \tr \Big( \frac{1}{4}F_{\mu \nu} F^{\mu \nu} + \frac{1}{2} \D_{\mu} \si \D^{\mu} \si + \frac{1}{2} (D + \frac{\si}{f})^2 \nonumber \\
& \ \ \ \ \ \ \ \ \ \ \ \ \ \ \ \ \ \ \ \ \ \ \ \ \ + \frac{i}{2}\olam \g^{\mu} \D_{\mu} \lam + \frac{i}{2} \olam [ \si, \lam ] - \frac{1}{4f} \olam \lam \Big) \nonumber \\
&=\int d^3 x \ \sqrt{g} \ \dl_{\oep} \dl_{\e} \ \tr (\frac{1}{2} \olam \lam - 2 D \si).
\end{align}
Locus that gives the bosonic part of $S_{YM}$ as zero is characterized by
\begin{equation}
F_{\mu \nu}= 0, \ \ \D_{\mu} \si = 0 , (D + \frac{\si}{f}) = 0
\end{equation}
which is equivalent to
\begin{equation}
A_{\mu}= 0, \ \  \si = \si_0 (\text{constant}) , D =- \frac{\si_0}{f}
\end{equation}
up to gauge transformation.
\subsection{Localization formula}
Let us define the supersymmetric Wilson loop as the usual one
\begin{equation}
W_S(R,C;A, \si)= \tr _R \mathcal{P} \exp \Big( \oint_C d \tau( i A_{\mu} \dot{x}^{\mu} + \si |\dot{x}| )  \Big) .
\end{equation}
The main statement of localization theorem is following.
The function
\begin{equation}
W_S (t) = \int \D A \ \D \lam \ \D \olam \ \D D \ \D \si \ e^{i\frac{k}{4\pi}S_{SCS}(A, \si, \lam, \olam, D)-t S_{YM}(A, \si, \lam, \olam, D)} W_S(R,C;A, \si)
\end{equation}
does not depend on $t$ if and only if 
\begin{equation}
\dl_{\oep} W_S(R,C;A, \si) = 0 \label{1/2BPS}
\end{equation}
is satisfied.

Therefore, if we satisfy this $1/2$ BPS condition, 
\begin{equation}
\lim_{t \to +0} W_S(t) = \lim_{t \to +\infty} W_S(t) \label{loc}
\end{equation}
is valid. 
Let us define this value as $W$.

Naively the left hand side of (\ref{loc}) looks to be level $k$ Chern-Simons theory, however when we consider the gauge group as $U(N)$ or $SU(N)$, it turns to be level $k-N$ Chern-Simons theory.
We will get back to this issue in section 4.
On the right hand side of (\ref{loc}), field configuration localize at the locus that makes the bosonic part of $S_{YM}$ zero.
Then, after performing the usual localization procedure, we get
\begin{align}
W= \lim_{t \to +\infty} W_S(t) 
&= \int_{\text{Cartan}} d \si_0 \ e^{i\frac{k}{4\pi}S_{SCS}(0, \si_0, 0, 0, -\frac{\si_0}{f})}W_S(R,C;0, \si_0) \times \mathcal{Z}_{\text{1-loop}}^{(HHL)} (\si_0) \nonumber \\
&=  \int_{\text{Cartan}} d \si_0 \ e^{-ik\pi l \tilde{l} \tr (\si_0^2)}\tr_R(e^{ \si_0 \oint_C d \tau |\dot{x}|}) \times \mathcal{Z}_{\text{1-loop}}^{(HHL)} (\si_0), \label{formula}
\end{align}
where
\begin{equation}
\mathcal{Z}_{\text{1-loop}}^{(HHL)} (\si_0) = \prod_{\alpha>0} \sinh (\pi l \alpha(\si_0)) \sinh(\pi \tilde{l} \alpha(\si_0)) 
\end{equation}
as discussed in \cite{Hama:2011ea}.
Here, $\alpha$ means root and $\alpha > 0$ means positive roots.
In general, we can insert not one but many 1/2 BPS Wilson loops into path integral.
Assume there are many 1/2 BPS contours denoted as $C_i \ (i=1,2,...,n)$, then we get
\begin{align}
& \int \D A \ \D \lam \ \D \olam \ \D D \ \D \si \ e^{i\frac{k}{4\pi}S_{SCS}(A, \si, \lam, \olam, D)-t S_{YM}(A, \si, \lam, \olam, D)} \prod_{i=1} ^{n} W_S(R_i,C_i;A, \si) \nonumber \\
&=  \int_{\text{Cartan}} d \si_0 \ e^{-ik\pi l \tilde{l} \tr (\si_0^2)} \prod_{i=1} ^{n}\tr_{R_i}(e^{ \si_0 \oint_{C_i} d \tau |\dot{x}|}) \times \mathcal{Z}_{\text{1-loop}}^{(HHL)} (\si_0) , \label{link}
\end{align}
where $R_i$ is a representation assigned with the loop $C_i$.
Let us define this value as $W_{12...n} $.

And as usual, let us define the partition function as
\begin{equation}
Z= \int_{\text{Cartan}} d \si_0 \ e^{-ik\pi l \tilde{l} \tr (\si_0^2)} \mathcal{Z}_{\text{1-loop}}^{(HHL)} (\si_0) .
\end{equation}
\newpage
\section{1/2 BPS Wilson loop and Seifert fibrations }\label{sec3}

\subsection{1/2 BPS condition}
In order to evaluate (\ref{formula}) or (\ref{link}), we have to know $\oint_C d \tau |\dot{x}|$ that is determined by 1/2 BPS condition (\ref{1/2BPS}).
It reduces to following condition
\begin{equation}
\dl_{\oep} W_S(R,C;A, \si) \propto \frac{1}{2} \oep(\g_{\mu} \dot{x}^{\mu} + |\dot{x}|) \lam=0.
\end{equation}
Therefore, we must solve following ODE
\begin{equation}
 \oep(\g_{\mu} \dot{x}^{\mu} + |\dot{x}|)=0.
\end{equation}
Substituting the explicit form of $\oep$ (\ref{killing}) and taking $|\dot{x}|=1$, this condition is equivalent to
\begin{eqnarray}
\dot{x}^{\mu} \frac{\pa}{\pa x^{\mu}} =
\left\{ \begin{array}{ll}
\frac{1}{l} \frac{\pa}{\pa \phi} - \frac{1}{\tilde{l}} \frac{\pa}{\pa \chi} & (\theta \neq  0, \frac{\pi}{2} ) \\
\frac{1}{l} \frac{\pa}{\pa \phi}  & (\theta =  0 ) \\
- \frac{1}{\tilde{l}} \frac{\pa}{\pa \chi} & (\theta =\frac{\pi}{2} ) \\
\end{array} \right. . \label{1/2BPSWl}
\end{eqnarray}
Let us investigate each situation in detail.
\subsection{The shape of each loop and Jones polynomials}
\begin{figure}[htbp]
 \begin{minipage}{0.5\hsize}
 \vspace{0.7cm}
  \begin{center}
   \includegraphics[width=30mm]{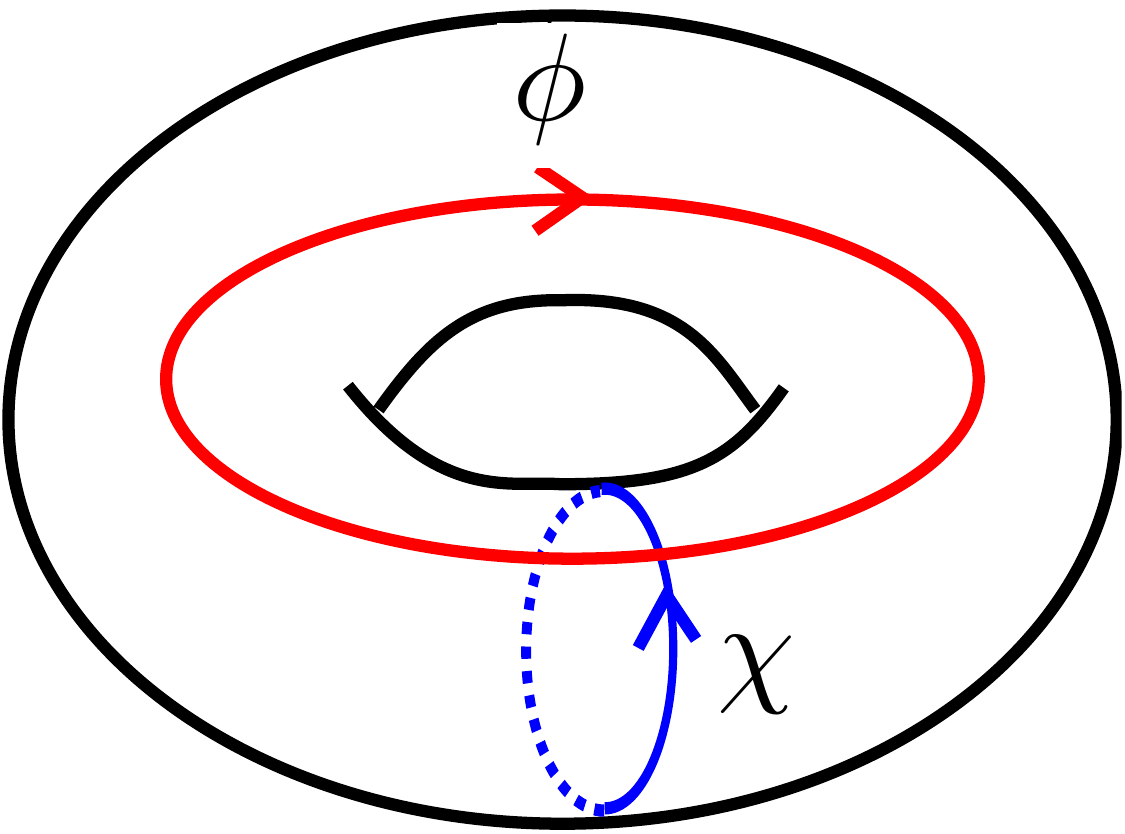}
  \end{center}
  \caption{$T^2$}
  \label{torus}
 \end{minipage}
 \begin{minipage}{0.5\hsize}
  \begin{center}
   \includegraphics[width=60mm]{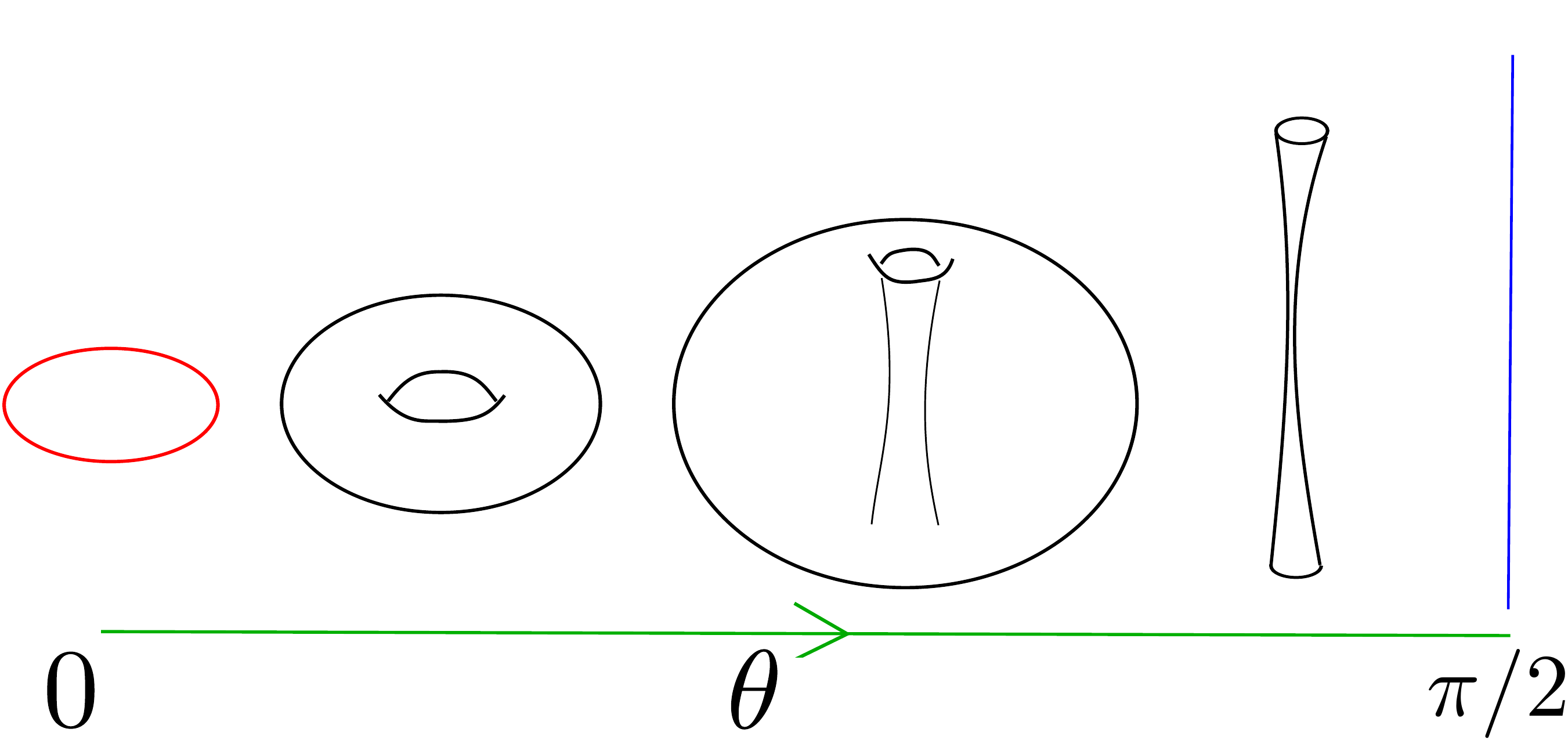}
  \end{center}
  \caption{$S^3$ as $R^3 \cup \{ \infty \}$}
  \label{theta}
 \end{minipage}
\end{figure}

First of all, we regard $S^3$ as one point compactified $R^3$.
Through this picture, we can visualize coordinates $\phi, \chi, \theta$.
$\phi$ and $\chi$ represent 2-dimensional torus $T^2$ (Figure \ref{torus}).

$\theta$ can be regarded as the size of this torus.
During $0 < \theta < \pi/2$, this torus changes its size as described in Figure \ref{theta}.
When $\theta=0$, $\chi$ shrinks to one point and torus reduces to a circle (red one in Figure \ref{theta}).
If we grow up this torus bigger and bigger, the size tends to infinity, and when we reach $\theta =\pi /2$, $\phi$ vanishes into far away.
Then torus reduces to a line (blue one in Figure \ref{theta}).
However we should regard it not a line but a circle because of the one point compactification.\\

\subsubsection*{1/2 BPS Wilson loop on $\theta \neq 0, \pi /2$}
\begin{center}
\begin{figure}[htbp]
  \begin{center}
   \includegraphics[width=70mm]{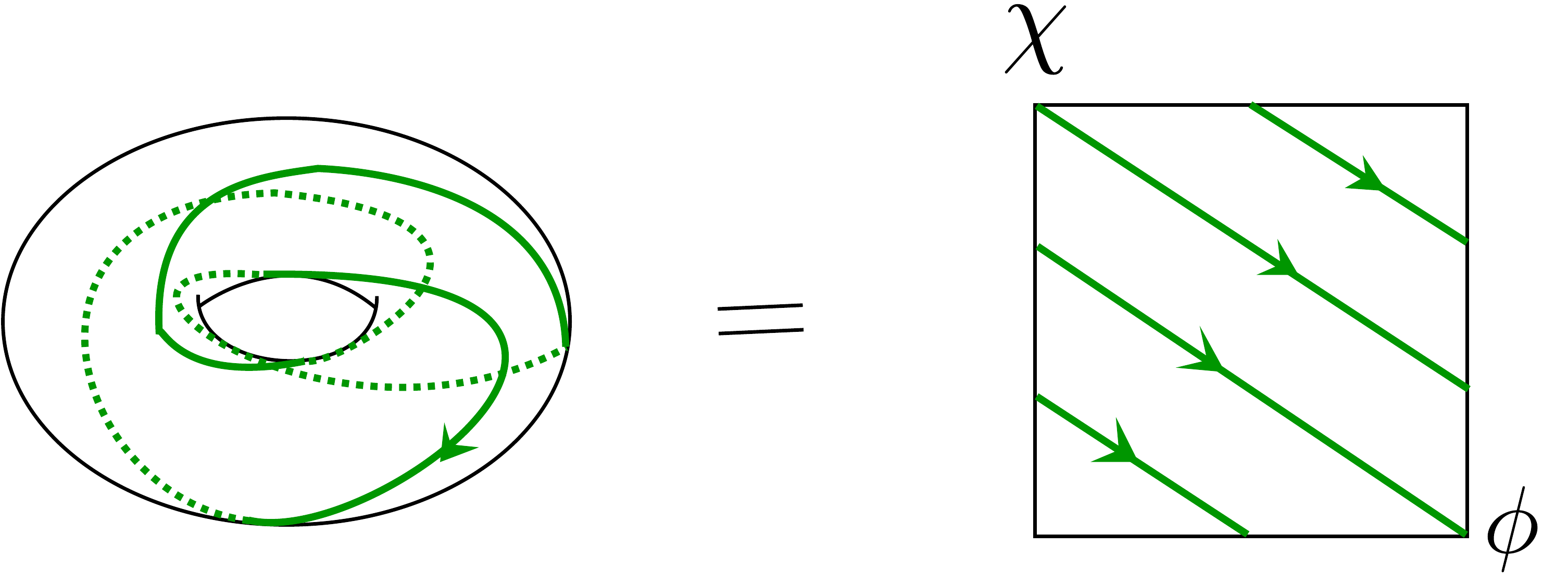}
  \end{center}
  \caption{Green line corresponds to (\ref{torusknot}).}
  \label{torus knot}
 \end{figure}
\end{center}

In this case, according to (\ref{1/2BPSWl}), we get 1/2 BPS Wilson loop as
\begin{equation}
\phi = - \frac{\tilde{l}}{l} \chi + \phi_0, \ \ \theta = \theta_0 \neq 0, \frac{\pi}{2} \label{torusknot}
\end{equation}
where $\phi_0, \theta_0$ are integration constants.
As shown in Figure \ref{torus knot}, this curve becomes closed loop if and only if $\tilde{l}/l$ is a rational number\footnote{If not, the integral curve cannot get back to the initial point, and wraps the torus densely. } .
We can get the length of this loop as
\begin{align}
\oint d\tau |\dot{x}| 
&= \oint d \tau = 2 \pi l \tilde{l}.
\end{align}
Therefore, in this case we get
\begin{equation}
W
=\int_{\text{Cartan}} d \si_0 \ e^{-ik\pi l \tilde{l} \tr (\si_0^2)}\tr_R(e^{2 \pi l \tilde{l} \si_0 }) \times \mathcal{Z}_{\text{1-loop}}^{(HHL)} (\si_0).
\end{equation}
As a consistency check, we calculate this integration.
For simplicity, we take gauge group as $U(2)$ and $R=2$.
The result is
\begin{equation}
\frac{W}{Z} =  q^{-(l+1)(\tilde{l}+1)/2} \frac{-1}{q^{1/2}-q^{-1/2}} (q^{l + \tilde{l}} + 1 -  q^{l+1} - q^{\tilde{l} +1}), \label{jones}
\end{equation} 
where
\begin{align}
&q = e^{\frac{2 \pi i}{k}} \label{q}.
\end{align}
This polynomial is well known $(l, \tilde{l})$-torus knot Jones polynomial up to extra phase factor $e^{-l \tilde{l}\frac{2 \pi i}{k}}$.
As pointed out in \cite{2011arXiv1101.3216W}, we must use normalized Jones polynomial so that the polynomial of trivial knot becomes $\frac{q-q^{-1}}{q^{1/2}-q^{-1/2}}$ .
After this renormalization procedure, by using the well-known formula of torus-knot \cite{Jones2005}, we get $(l, \tilde{l})$-torus knot Jones polynomial
\begin{equation}
J_{l,\tilde{l}} (q) = q^{l\tilde{l}} q^{-(l+1)(\tilde{l}+1)/2} \frac{-1}{q^{1/2}-q^{-1/2}} (q^{l + \tilde{l}} + 1 -  q^{l+1} - q^{\tilde{l} +1}). \label{Jll}
\end{equation}
\newpage
\subsubsection*{1/2 BPS Wilson loop on $\theta=0$}
\begin{center}
\begin{figure}[htbp]
  \begin{center}
   \includegraphics[width=60mm]{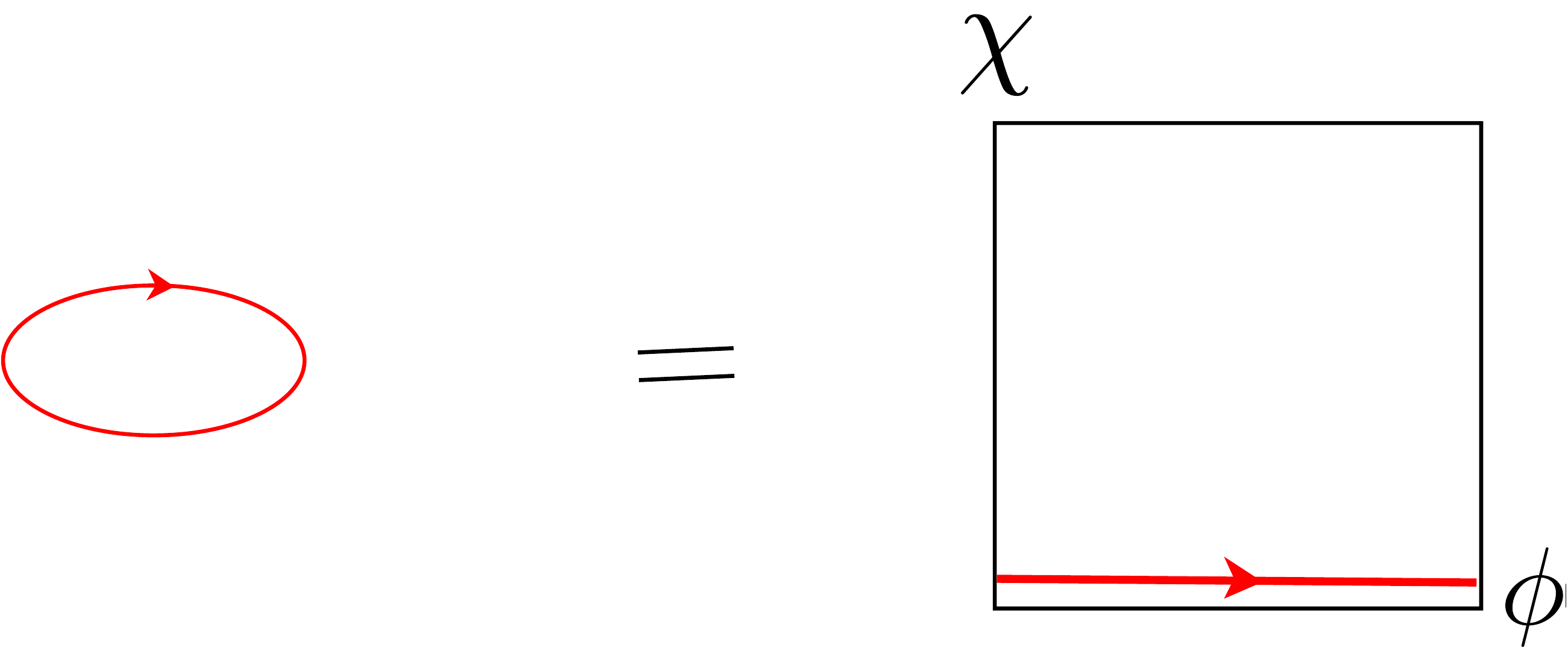}
  \end{center}
  \caption{1/2 BPS Wilson loop on $\theta = 0$}
  \label{zero}
 \end{figure}
\end{center}
In this case (figure \ref{zero}), (\ref{1/2BPSWl}) says the length is 
\begin{align}
\oint d\tau |\dot{x}| 
=&\oint d \tau = 2 \pi l .
\end{align}
Therefore (\ref{formula}) turns to be
\begin{equation}
W
=  \int_{\text{Cartan}} d \si_0 \ e^{-ik\pi l \tilde{l} \tr (\si_0^2)}\tr_R(e^{ 2 \pi l \si_0 }) \times \mathcal{Z}_{\text{1-loop}}^{(HHL)} (\si_0).
\end{equation}
In the case of $U(2)$ gauge theory and $R=2$, we get
\begin{equation}
\frac{W }{Z} = e^{- \frac{l}{\tilde{l}} \frac{2 \pi i}{k}} \frac{q- q^{-1}}{q^{1/2}-q^{-1/2}}. \label{south}
\end{equation}
\subsubsection*{1/2 BPS Wilson loop on $\theta=\pi /2$}
\begin{center}
\begin{figure}[htbp]
  \begin{center}
   \includegraphics[width=55mm]{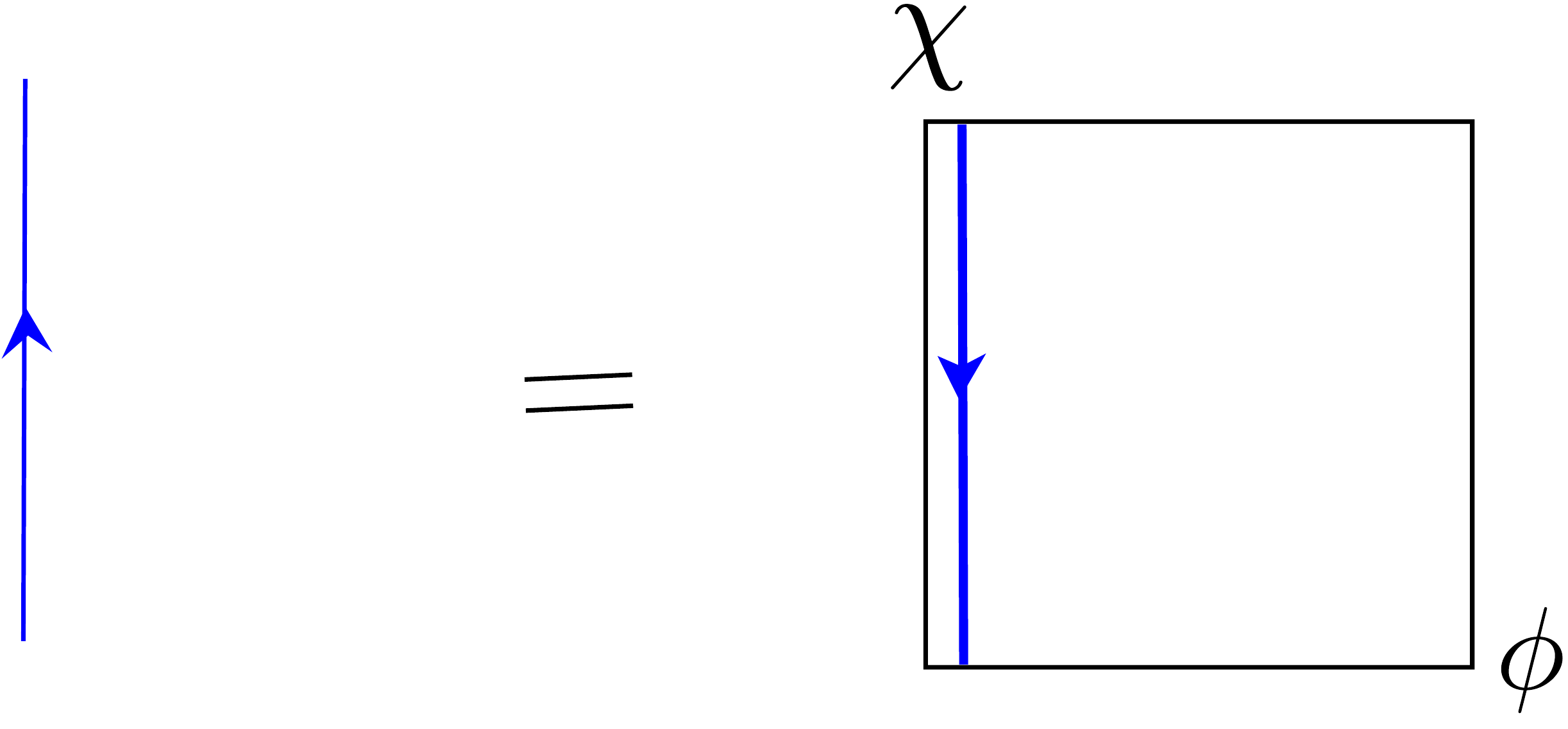}
  \end{center}
  \caption{1/2 BPS Wilson loop on $\theta = \pi/2$}
  \label{pi}
 \end{figure}
\end{center}
In this case (figure \ref{pi}), (\ref{1/2BPSWl}) says the length is
\begin{align}
\oint d\tau |\dot{x}| 
=&\oint d \tau = 2 \pi \tilde{l} .
\end{align}
Therefore (\ref{formula}) turns to be
\begin{equation}
W
=  \int_{\text{Cartan}} d \si_0 \ e^{-ik\pi l \tilde{l} \tr (\si_0^2)}\tr_R(e^{ 2 \pi \tilde{l} \si_0 }) \times \mathcal{Z}_{\text{1-loop}}^{(HHL)} (\si_0).
\end{equation}
In the case of $U(2)$ gauge theory and $R=2$, we get
\begin{equation}
\frac{W }{Z} = e^{- \frac{\tilde{l}}{l} \frac{2 \pi i}{k}} \frac{q- q^{-1}}{q^{1/2}-q^{-1/2}}.\label{north}
\end{equation}

(\ref{south}) and (\ref{north}) are trivial knot Jones polynomials up to extra phase factors $e^{- \frac{l}{\tilde{l}} \frac{2 \pi i}{k}},e^{- \frac{\tilde{l}}{l} \frac{2 \pi i}{k}}$ respectively.

\subsection{Hopf link (on north pole and south pole) and Seifert fibration}
\begin{figure}[htbp]
 \begin{minipage}{0.5\hsize}
  \begin{center}
   \includegraphics[width=15mm]{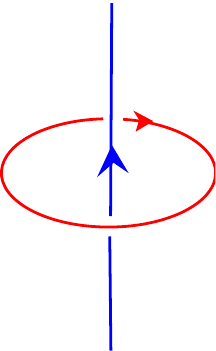}
  \end{center}
  \caption{Left handed Hopf link}
  \label{hopf}
 \end{minipage}
 \begin{minipage}{0.5\hsize}
  \begin{center}
   \includegraphics[width=20mm]{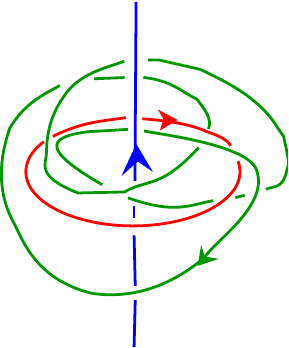}
  \end{center}
  \caption{Seifert fibration}
  \label{seifert}
 \end{minipage}
\end{figure}

As discussed above, we can also evaluate correlation function of many 1/2 BPS Wilson loops.
As the simplest example, we put two Wilson loops on the north pole ($\theta=\pi/2$) and the south pole ($\theta=0$).
These loops $C_{\theta=\pi/2}$ and $C_{\theta=0}$ form so-called Hopf link (Figure \ref{hopf}).

The result is
\begin{align}
W_{\theta=\pi/2, \theta=0}=\int_{\text{Cartan}} d \si_0 \ e^{-ik\pi l \tilde{l} \tr (\si_0^2)} \tr_{R_{\theta=\pi/2}}(e^{2 \pi \tilde{l} \si_0 }) \tr_{R_{\theta=0}}(e^{ 2 \pi l  \si_0}) \times \mathcal{Z}_{\text{1-loop}}^{(HHL)} (\si_0).
\end{align}
We can get the simplest result by taking $R_{\theta=\pi/2}=R_{\theta=0}=2$.
Then, this reduces to the form
\begin{equation}
\frac{W_{\theta=\pi/2, \theta=0}}{Z} =e^{-(2+\frac{\tilde{l}}{l}+\frac{l}{\tilde{l}})\frac{2 \pi i}{k}} (q^3+q^2+q+1).
\end{equation}
In this case, we have extra phase factor as $e^{-(2+\frac{\tilde{l}}{l}+\frac{l}{\tilde{l}})\frac{2 \pi i}{k}}$.

When we gather all of 1/2 BPS Wilson loops, they wrap whole of the $S^3$.
In addition, this wrapping means that we can separate $S^3$ into each $U(1)$ equivalent classes.
And because of $S^3 = SU(2)$ and $S^3/U(1) = SU(2)/U(1) = S^2$ we get nontrivial fibration structure $U(1) \to S^3 \to S^2$.
This fibration is called Seifert fibration (Figure \ref{seifert}).

\section{Comments on anomalies}\label{Sec4}

\subsection{Parity anomaly}
\begin{center}
\begin{figure}[htbp]
  \begin{center}
   \includegraphics[width=50mm]{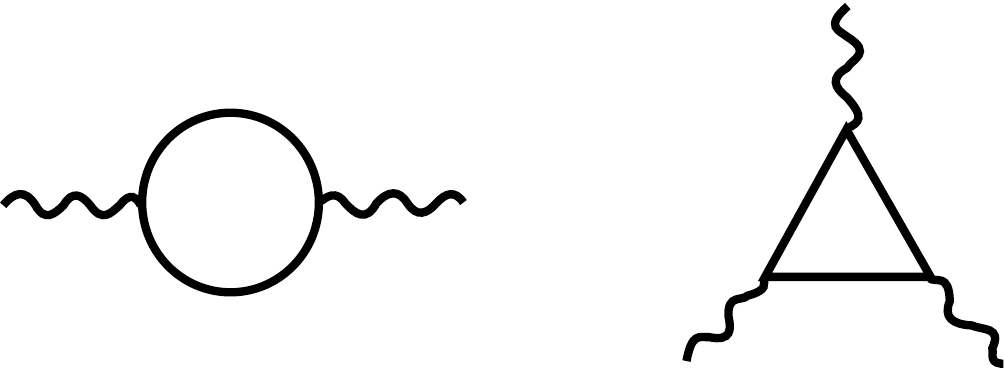}
  \end{center}
  \caption{Leading contributions}
  \label{1loop}
 \end{figure}
\end{center}

One may naively expect that the corresponding result on pure Chern-Simons side is level k, because
\begin{equation}
Z(t) \Big|_{t=0} = (\text{constant}) \times \int \D A \ \exp{ \Big( \frac{ik}{4\pi}  \int d^3 x \  \e ^{\mu \nu \lam} \tr (A_{\mu} \pa_{\nu} A_{\lam} - \frac{2i}{3}A_{\mu} A_{\nu} A_{\lam} ) \Big)} .
\end{equation}
However, if it is true, the definition $q = e^{\frac{2 \pi i}{k}}$ in our paper conflicts with that of \cite{Witten:1988hf}.
Here, it is important to be careful with the order of taking limit.
As pointed out in \cite{Kao:1995gf, Kapustin:2010mh}, we have to perform path integral first, then take limit $t \to +0$.
In this procedure, we encounter following effective action that comes from integration with gaugino\footnote{We would like to thank Y. Hosotani and T. Onogi who suggested this idea.}
\begin{equation}
\int \D \lam \ \D \olam \ e^{\text{gaugino term in } (i\frac{k}{4\pi}S_{SCS} - tS_{YM})} = e^{\Gamma(A,t)}.
\end{equation}
 Performing usual perturbative expansion, the leading terms come from 1-loop diagrams as shown in Figure \ref{1loop}.
 This phenomenon is well known in the context of parity anomaly \cite{Niemi:1983rq, Redlich:1983dv}.
When we take the gauge group as $U(N)$ or $SU(N)$, the result is level $(-N)$ pure Chern-Simons theory
 \begin{equation}
\Gamma(A,t) \to \frac{-iN}{4\pi}  \int d^3 x \  \e ^{\mu \nu \lam} \tr (A_{\mu} \pa_{\nu} A_{\lam} - \frac{2i}{3}A_{\mu} A_{\nu} A_{\lam} ) , \ \ (t \to +0).
\end{equation}
Therefore, we can conclude
\begin{equation}
\lim_{t \to +0} Z(t) =(\text{constant}) \times \int \D A \ \exp{ \Big( \frac{i(k-N)}{4\pi}  \int d^3 x \  \e ^{\mu \nu \lam} \tr (A_{\mu} \pa_{\nu} A_{\lam} - \frac{2i}{3}A_{\mu} A_{\nu} A_{\lam} ) \Big)} .
\end{equation}
\subsection{Framing anomaly}
As commented in \cite{Kapustin:2009kz}, we have the framing anomaly as 
\begin{align}
\frac{W_{12...n}}{Z} = e^{\sum_{i,j =1}^{n} \text{lk}(C_i,C_j) \frac{2 \pi i}{k}} \times \text{Jones polynomial of link}(C_1,C_2,...,C_n),
\end{align}
where $\text{lk}(C_i,C_j)$ means linking number between $C_i$ and $C_j$\footnote{The framing anomaly of three manifold itself is cancelled because of dividing $W_{12,...,n}$ by $Z$.}.
As we can see, even if there is only one knot, we have the phase $e^{\text{lk}(C,C) \frac{2\pi i}{k}}$.
$\text{lk}(C,C)$ is called self linking number. This value is UV divergent, therefore we must regularize it by using point-splitting regularization \cite{Witten:1988hf}.
Here, we evaluated 1/2 BPS Wilson loops. In our calculations, this regularization scheme is automatically selected in order to maintain the 1/2 BPS condition \cite{Kapustin:2009kz}.
In other words, we calculated $\text{lk}(C,C)$ as 
\begin{equation}
\text{lk}(\text{one 1/2 BPS loop },\text{another 1/2 BPS loop}).
\end{equation}
We would like to call another 1/2 BPS Wilson loop as splitting loop.

\subsubsection*{Torus knot case}
In our orientation, all of 1/2 BPS Wilson loops are constructed as left-handed manner.
Therefore, all linking numbers are negative.
In general, left-handed $(l, \tilde{l})$-torus knot has linking number as $-l \tilde{l}$.
So, our result must match with
\begin{equation}
e^{-l \tilde{l}\frac{2 \pi i}{k}} \times J_{l,\tilde{l}} (q).
\end{equation}
According to (\ref{jones}) and (\ref{Jll}), this is satisfied.

\subsubsection*{Trivial knot case}
According to (\ref{south}) and (\ref{north}), we seem to have rational numbers as these linking numbers as $ -\frac{l}{\tilde{l}}$ and $-\frac{\tilde{l}}{l}$.
What is the meaning of rational linking number?

1/2 BPS Wilson loop near the trivial knot on $\theta = 0$ is $(l,\tilde{l})$-torus knot.
Linking number between these two knots is $-l$ because $(l, \tilde{l})$-torus knot wraps $l$ times around the trivial knot on $\theta = 0$ in the left handed manner.

As we calculated in section 3, these loops have their length as $2 \pi l$ and $2 \pi l \tilde{l}$.
Here the ratio of these length is $2\pi l / 2 \pi l \tilde{l} = 1/\tilde{l}$.
It means that during a test particle wraps the trivial knot on $\theta =0$, another test particle on torus knot cannot travel whole of the loop but $1/\tilde{l}$ of it.
Then, $1/ \tilde{l}$ part of torus knot wraps $-l/\tilde{l}$ times around centered trivial knot, i.e.
\begin{equation}
\text{lk}(C_{\theta=0},C_{\theta=0})=-l/\tilde{l}.
\end{equation}
As same,
\begin{equation}
\text{lk}(C_{\theta=\pi/2},C_{\theta=\pi/2})=-\tilde{l}/l.
\end{equation}
This interpretation does not depend on the choice of splitting 1/2 BPS Wilson loop.

\subsubsection*{Hopf link case}
The phase $e^{-(2+\frac{\tilde{l}}{l}+\frac{l}{\tilde{l}})\frac{2 \pi i}{k}}$ can be interpreted as
\begin{align}
&\sum_{i,j =1}^{n} \text{lk}(C_i,C_j) \frac{2 \pi i}{k} \nonumber \\
&= (\text{lk}(C_{\theta=\pi/2},C_{\theta=\pi/2})+\text{lk}(C_{\theta=\pi/2},C_{\theta=0})+\text{lk}(C_{\theta=0},C_{\theta=\pi/2})+\text{lk}(C_{\theta=0},C_{\theta=0}))\frac{2 \pi i}{k} \nonumber \\
&= (- \frac{\tilde{l}}{l}-1-1-\frac{l}{\tilde{l}})\frac{2 \pi i}{k} =  -(2+ \frac{\tilde{l}}{l}+\frac{l}{\tilde{l}})\frac{2 \pi i}{k}.
\end{align}

\section{Discussion}
We checked the consistency between localization techniques and well known exact results of Chern-Simons theory
and found complete equivalence as expected.
Another squashing is discussed in \cite{Imamura:2011wg} and
localization on other three-dimensional manifolds is discussed by \cite{Gang:2009wy} on lens spaces, \cite{2011JHEP...08..008K} on general Seifert manifolds by performing topological twist on $\mathcal{N}=2$ vector multiplet.
It might be possible to construct more generic knot matrix model by using these modified localization techniques.

Our result (\ref{formula}) is the well known knot matrix model \cite{Beasley:2009mb,2011JHEP...08..008K}.
However our derivation is simpler and more comprehensive because we only use supersymmetry.

Originally, the squashing techniques were developed in order to study domain walls in four-dimensional $\mathcal{N}=2$ gauge theories \cite{Drukker:2010jp, Hosomichi:2010vh}.
It may be interesting to apply our results to these studies.

\vspace{0.2cm}
\noindent
\section*{Acknowledgements}
I would like to thank S. Yamaguchi, Y. Hosotani, T. Onogi and Y. Tachikawa, V. Pestun, D. Gang for valuable discussions and comments.
And I would also like to thank H. Tanida, T. Nishinaka, K.Oda for encouragement, and W. Naylor for a careful reading of this manuscript and useful comments.


\providecommand{\href}[2]{#2}\begingroup\raggedright\endgroup

\end{document}